\documentstyle[epsf,epsfig,here,12pt]{article}  
\begin{document}

\newcommand{\HR}{\rule{1em}{0.4pt}}
\newcommand{\rhobar}{\overline{\rho}}
\newcommand{\etabar}{\overline{\eta}}
\newcommand{\epsilonk}{\left|\epsilon_K \right|}
\newcommand{\vubovcb}{\left | \frac{V_{ub}}{V_{cb}} \right |}
\newcommand{\vtdovts}{\left | \frac{V_{td}}{V_{ts}} \right |}
\newcommand{\dmd}{\Delta m_d}
\newcommand{\dms}{\Delta m_s}
\newcommand{\pr}{{\rm P.R.}}
\newcommand{\Ds}{{\rm D}_s^+}
\newcommand{\Dp}{{\rm D}^+}
\newcommand{\Do}{{\rm D}^0}
\newcommand{\piss}{\pi^{\ast \ast}}
\newcommand{\pis}{\pi^{\ast}}
\newcommand{\bbar}{\overline{b}}
\newcommand{\cbar}{\overline{c}}
\newcommand{\Dstar}{{\rm D}^{\ast}}
\newcommand{\Dstars}{{\rm D}^{\ast +}_s}
\newcommand{\Dstaro}{{\rm D}^{\ast 0}}
\newcommand{\Dstarstar}{{\rm D}^{\ast \ast}}
\newcommand{\Dbar}{\overline{{\rm D}}}
\newcommand{\Bbar}{\overline{{\rm B}}}
\newcommand{\Bsbar}{\overline{{\rm B}^0_s}}
\newcommand{\Lcbar}{\overline{\Lambda^+_c}}
\newcommand{\nubar}{\overline{\nu_{\ell}}}
\newcommand{\tautaubar}{\tau \overline{\tau}}
\newcommand{\Vcb}{\left | {\rm V}_{cb} \right |}
\newcommand{\Vub}{\left | {\rm V}_{ub} \right |}
\newcommand{\Vtd}{\left | {\rm V}_{td} \right |}
\newcommand{\Vts}{\left | {\rm V}_{ts} \right |}
\newcommand{\fleisher}{\frac{BR({\rm B}^0~(\overline{{\rm B}^0}) \rightarrow \pi^{\pm} {\rm K}^{\mp})}
{BR({\rm B}^{\pm} \rightarrow \pi^{\pm} {\rm K}^0)}}
\newcommand{\bptre}{\rm b^{+}_{3}}
\newcommand{\bp}{\rm b^{+}_{1}}
\newcommand{\bo}{\rm b^0}
\newcommand{\bos}{\rm b^0_s}
\newcommand{\bss}{\rm b^s_s}
\newcommand{\qq}{\rm q \overline{q}}
\newcommand{\cc}{\rm c \overline{c}}
\newcommand{\BsDmX}{{B_{s}^{0}} \rightarrow D \mu X}
\newcommand{\BsDsm}{{B_{s}^{0}} \rightarrow D_{s} \mu X}
\newcommand{\BsDsX}{{B_{s}^{0}} \rightarrow D_{s} X}
\newcommand{\BDsX}{B \rightarrow D_{s} X}
\newcommand{\BDomX}{B \rightarrow D^{0} \mu X}
\newcommand{\BDpmX}{B \rightarrow D^{+} \mu X}
\newcommand{\Dsfmn}{D_{s} \rightarrow \phi \mu \nu}
\newcommand{\Dsfipi}{D_{s} \rightarrow \phi \pi}
\newcommand{\DsfX}{D_{s} \rightarrow \phi X}
\newcommand{\DpfX}{D^{+} \rightarrow \phi X}
\newcommand{\DofX}{D^{0} \rightarrow \phi X}
\newcommand{\DfX}{D \rightarrow \phi X}
\newcommand{\DsD}{B \rightarrow D_{s} D}
\newcommand{\DsmX}{D_{s} \rightarrow \mu X}
\newcommand{\DmX}{D \rightarrow \mu X}
\newcommand{\Zbb}{Z^{0} \rightarrow \rm b \overline{b}}
\newcommand{\Zcc}{Z^{0} \rightarrow \rm c \overline{c}}
\newcommand{\Rbb}{\frac{\Gamma_{Z^0 \rightarrow \rm b \overline{b}}}
{\Gamma_{Z^0 \rightarrow Hadrons}}}
\newcommand{\Rcc}{\frac{\Gamma_{Z^0 \rightarrow \rm c \overline{c}}}
{\Gamma_{Z^0 \rightarrow Hadrons}}}
\newcommand{\bb}{\rm b \overline{b}}
\newcommand{\str}{\rm s \overline{s}}
\newcommand{\Bs}{\rm{B^0_s}}
\newcommand{\Bsb}{\overline{\rm{B^0_s}}}
\newcommand{\Bp}{\rm{B^{+}}}
\newcommand{\Bm}{\rm{B^{-}}}
\newcommand{\Bo}{\rm{B^{0}}}
\newcommand{\Bd}{\rm{B^{0}_{d}}}
\newcommand{\Bdb}{\overline{\rm{B^{0}_{d}}}}
\newcommand{\Lb}{\Lambda^0_b}
\newcommand{\Lbb}{\overline{\Lambda^0_b}}
\newcommand{\Kstar}{\rm{K^{\star 0}}}
\newcommand{\phim}{\rm{\phi}}
\newcommand{\Dsp}{\mbox{D}_s^+}
\newcommand{\Dn}{\mbox{D}^0}
\newcommand{\Dsb}{\overline{\mbox{D}_s}}
\newcommand{\Dm}{\mbox{D}^-}
\newcommand{\Dnb}{\overline{\mbox{D}^0}}
\newcommand{\Lc}{\Lambda_c}
\newcommand{\Lcb}{\overline{\Lambda_c}}
\newcommand{\Dstarp}{\mbox{D}^{\ast +}}
\newcommand{\Dstarm}{\mbox{D}^{\ast -}}
\newcommand{\Dsstarp}{\mbox{D}_s^{\ast +}}
\newcommand{\Km}{\mbox{K}^-}
\newcommand{\Pb}{P_{b-baryon}}
\newcommand{\KKpi}{\rm{ K K \pi }}
\newcommand{\GeV}{\rm{GeV}}
\newcommand{\MeV}{\rm{MeV}}
\newcommand{\nb}{\rm{nb}}
\newcommand{\Zzero}{{\rm Z}^0}
\newcommand{\MZ}{\rm{M_Z}}
\newcommand{\MW}{\rm{M_W}}
\newcommand{\GF}{\rm{G_F}}
\newcommand{\Gm}{\rm{G_{\mu}}}
\newcommand{\MH}{\rm{M_H}}
\newcommand{\MT}{\rm{m_{top}}}
\newcommand{\GZ}{\Gamma_{\rm Z}}
\newcommand{\Afb}{\rm{A_{FB}}}
\newcommand{\Afbs}{\rm{A_{FB}^{s}}}
\newcommand{\sigmaf}{\sigma_{\rm{F}}}
\newcommand{\sigmab}{\sigma_{\rm{B}}}
\newcommand{\NF}{\rm{N_{F}}}
\newcommand{\NB}{\rm{N_{B}}}
\newcommand{\Nnu}{\rm{N_{\nu}}}
\newcommand{\RZ}{\rm{R_Z}}
\newcommand{\rhob}{\rho_{eff}}
\newcommand{\Gammanz}{\rm{\Gamma_{Z}^{new}}}
\newcommand{\Gammani}{\rm{\Gamma_{inv}^{new}}}
\newcommand{\Gammasz}{\rm{\Gamma_{Z}^{SM}}}
\newcommand{\Gammasi}{\rm{\Gamma_{inv}^{SM}}}
\newcommand{\Gammaxz}{\rm{\Gamma_{Z}^{exp}}}
\newcommand{\Gammaxi}{\rm{\Gamma_{inv}^{exp}}}
\newcommand{\rhoZ}{\rho_{\rm Z}}
\newcommand{\thw}{\theta_{\rm W}}
\newcommand{\swsq}{\sin^2\!\thw}
\newcommand{\swsqmsb}{\sin^2\!\theta_{\rm W}^{\overline{\rm MS}}}
\newcommand{\swsqbar}{\sin^2\!\overline{\theta}_{\rm W}}
\newcommand{\cwsqbar}{\cos^2\!\overline{\theta}_{\rm W}}
\newcommand{\swsqb}{\sin^2\!\theta^{eff}_{\rm W}}
\newcommand{\ee}{{e^+e^-}}
\newcommand{\eeX}{{e^+e^-X}}
\newcommand{\gaga}{{\gamma\gamma}}
\newcommand{\mumu}{\ifmmode {\mu^+\mu^-} \else ${\mu^+\mu^-} $ \fi}
\newcommand{\eeg}{{e^+e^-\gamma}}
\newcommand{\mumug}{{\mu^+\mu^-\gamma}}
\newcommand{\tautau}{{\tau^+\tau^-}}
\newcommand{\qqb}{{q\bar{q}}}
\newcommand{\eegg}{e^+e^-\rightarrow \gamma\gamma}
\newcommand{\eeggg}{e^+e^-\rightarrow \gamma\gamma\gamma}
\newcommand{\eeee}{e^+e^-\rightarrow e^+e^-}
\newcommand{\eeeeee}{e^+e^-\rightarrow e^+e^-e^+e^-}
\newcommand{\eeeeg}{e^+e^-\rightarrow e^+e^-(\gamma)}
\newcommand{\eeeegg}{e^+e^-\rightarrow e^+e^-\gamma\gamma}
\newcommand{\eeeg}{e^+e^-\rightarrow (e^+)e^-\gamma}
\newcommand{\eemumu}{e^+e^-\rightarrow \mu^+\mu^-}
\newcommand{\eetautau}{e^+e^-\rightarrow \tau^+\tau^-}
\newcommand{\eehad}{e^+e^-\rightarrow {\rm hadrons}}
\newcommand{\eettg}{e^+e^-\rightarrow \tau^+\tau^-\gamma}
\newcommand{\eell}{e^+e^-\rightarrow l^+l^-}
\newcommand{\Ztopig}{{\rm Z}^0\rightarrow \pi^0\gamma}
\newcommand{\Ztogg}{{\rm Z}^0\rightarrow \gamma\gamma}
\newcommand{\Ztoee}{{\rm Z}^0\rightarrow e^+e^-}
\newcommand{\Ztoggg}{{\rm Z}^0\rightarrow \gamma\gamma\gamma}
\newcommand{\Ztomumu}{{\rm Z}^0\rightarrow \mu^+\mu^-}
\newcommand{\Ztotautau}{{\rm Z}^0\rightarrow \tau^+\tau^-}
\newcommand{\Ztoll}{{\rm Z}^0\rightarrow l^+l^-}
\newcommand{\Ztocc}{{\rm Z^0\rightarrow c \bar c}}
\newcommand{\Lamp}{\Lambda_{+}}
\newcommand{\Lamm}{\Lambda_{-}}
\newcommand{\Pt}{\rm P_{t}}
\newcommand{\Gee}{\Gamma_{ee}}
\newcommand{\Gpig}{\Gamma_{\pi^0\gamma}}
\newcommand{\Ggg}{\Gamma_{\gamma\gamma}}
\newcommand{\Gggg}{\Gamma_{\gamma\gamma\gamma}}
\newcommand{\Gmumu}{\Gamma_{\mu\mu}}
\newcommand{\Gtautau}{\Gamma_{\tau\tau}}
\newcommand{\Ginv}{\Gamma_{\rm inv}}
\newcommand{\Ghad}{\Gamma_{\rm had}}
\newcommand{\Gnu}{\Gamma_{\nu}}
\newcommand{\GnuSM}{\Gamma_{\nu}^{\rm SM}}
\newcommand{\Gll}{\Gamma_{l^+l^-}}
\newcommand{\Gff}{\Gamma_{f\overline{f}}}
\newcommand{\Gtot}{\Gamma_{\rm tot}}
\newcommand{\Rb}{\mbox{R}_b}
\newcommand{\Rc}{\mbox{R}_c}
\newcommand{\al}{a_l}
\newcommand{\vl}{v_l}
\newcommand{\af}{a_f}
\newcommand{\vf}{v_f}
\newcommand{\ael}{a_e}
\newcommand{\ve}{v_e}
\newcommand{\amu}{a_\mu}
\newcommand{\vmu}{v_\mu}
\newcommand{\atau}{a_\tau}
\newcommand{\vtau}{v_\tau}
\newcommand{\ahatl}{\hat{a}_l}
\newcommand{\vhatl}{\hat{v}_l}
\newcommand{\ahate}{\hat{a}_e}
\newcommand{\vhate}{\hat{v}_e}
\newcommand{\ahatmu}{\hat{a}_\mu}
\newcommand{\vhatmu}{\hat{v}_\mu}
\newcommand{\ahattau}{\hat{a}_\tau}
\newcommand{\vhattau}{\hat{v}_\tau}
\newcommand{\vtildel}{\tilde{\rm v}_l}
\newcommand{\avsq}{\ahatl^2\vhatl^2}
\newcommand{\Ahatl}{\hat{A}_l}
\newcommand{\Vhatl}{\hat{V}_l}
\newcommand{\Afer}{A_f}
\newcommand{\Ael}{A_e}
\newcommand{\Aferb}{\bar{A_f}}
\newcommand{\Aelb}{\bar{A_e}}
\newcommand{\AVsq}{\Ahatl^2\Vhatl^2}
\newcommand{\Iwk}{I_{3l}}
\newcommand{\Qch}{|Q_{l}|}
\newcommand{\roots}{\sqrt{s}}
\newcommand{\pT}{p_{\rm T}}
\newcommand{\mt}{m_t}
\newcommand{\Rechi}{{\rm Re} \left\{ \chi (s) \right\}}
\newcommand{\up}{^}
\newcommand{\abscosthe}{|cos\theta|}
\newcommand{\dsum}{\Sigma |d_\circ|}
\newcommand{\zsum}{\Sigma z_\circ}
\newcommand{\sint}{\mbox{$\sin\theta$}}
\newcommand{\cost}{\mbox{$\cos\theta$}}
\newcommand{\mcost}{|\cos\theta|}
\newcommand{\epair}{\mbox{$e^{+}e^{-}$}}
\newcommand{\mupair}{\mbox{$\mu^{+}\mu^{-}$}}
\newcommand{\taupair}{\mbox{$\tau^{+}\tau^{-}$}}
\newcommand{\gamgam}{\mbox{$e^{+}e^{-}\rightarrow e^{+}e^{-}\mu^{+}\mu^{-}$}}
\newcommand{\fullskip}{\vskip 16cm}
\newcommand{\halfskip}{\vskip  8cm}
\newcommand{\quarskip}{\vskip  6cm}
\newcommand{\abitskip}{\vskip 0.5cm}
\newcommand{\ba}{\begin{array}}
\newcommand{\ea}{\end{array}}
\newcommand{\bc}{\begin{center}}
\newcommand{\ec}{\end{center}}
\newcommand{\be}{\begin{eqnarray}}
\newcommand{\eeq}{\end{eqnarray}}
\newcommand{\bes}{\begin{eqnarray*}}
\newcommand{\ees}{\end{eqnarray*}}
\newcommand{\Kz}{\ifmmode {\rm K^0_s} \else ${\rm K^0_s} $ \fi}
\newcommand{\Zz}{\ifmmode {\rm Z^0} \else ${\rm Z^0 } $ \fi}
\newcommand{\qqbar}{\ifmmode {\rm q\bar{q}} \else ${\rm q\bar{q}} $ \fi}
\newcommand{\ccbar}{\ifmmode {\rm c\bar{c}} \else ${\rm c\bar{c}} $ \fi}
\newcommand{\bbbar}{\ifmmode {\rm b\bar{b}} \else ${\rm b\bar{b}} $ \fi}
\newcommand{\xxbar}{\ifmmode {\rm x\bar{x}} \else ${\rm x\bar{x}} $ \fi}
\newcommand{\rphi}{\ifmmode {\rm R\phi} \else ${\rm R\phi} $ \fi}
\newcommand{\bt}{\begin{tabular}}
\newcommand{\et}{\end{tabular}}
\renewcommand{\arraystretch}{1.2}


\pagenumbering{arabic}

\vspace*{-1.8cm}
{\flushright{\bf LAL 99-03}\\
\vspace*{-0.5cm}
\flushright{March 1999}\\
\hspace*{11.5cm}{\small DELPHI 99-27 CONF 226}}\\


\vskip 7.5 cm

\begin{center}
{\bf\LARGE  Constraints on the parameters of the $CKM$ matrix}\\ 
\vspace*{0.5cm}
{\bf\LARGE  by End 1998}
\end{center}

\vskip 1.5 cm
\begin{center}
{\bf\Large  F. Parodi$^{(a)}$, P. Roudeau$^{(b)}$ and A. Stocchi$^{(b)}$}
\end{center}

\vspace*{4cm}

\noindent
{\bf $^{(a)}$ Dipartimento di Fisica, Universit\'a di Genova and INFN}\\
\hspace*{0.5cm}{\small Via Dodecaneso 33, 16146 Genova, Italy}

\vspace*{0.3cm}

\noindent
{\bf $^{(b)}$ Laboratoire de l'Acc\'el\'erateur Lin\'eaire}\\
\hspace*{0.5cm}{\small IN2P3-CNRS et Universit\'e de Paris-Sud, BP 34, F-91898 Orsay Cedex}
\newpage
\vspace*{-1.8cm}
{\flushright{\bf LAL 99-03}\\
\vspace*{-0.5cm}
\flushright{March 1999}\\
\hspace*{11.5cm}{\small DELPHI 99-27 CONF 226}}\\
\vspace*{-0.5cm}
\bc
\Large 
{\bf Constraints on the parameters of the ${CKM}$ matrix by End 1998. } \\
\vspace*{1.0cm}
\normalsize { {\bf Fabrizio Parodi$^{(a)}$, Patrick Roudeau$^{(b)}$ and Achille Stocchi$^{(b)}$ } }
\vspace{0.5 truecm}
\par {\it $^{(a)}$ Dipartimento di Fisica, Universit\'a di Genova and INFN, Via Dodecaneso 33, IT-16146 Genova, Italy \\
$^{(b)}$ Laboratoire de l'Acc\'elerateur Lin\'eaire (LAL), IN2P3-CNRS and Universit\'e de Paris-Sud, B.P. 34-91898 Orsay Cedex, 
France }

\vspace*{0.8cm}
\ec

\begin{abstract}
\noindent
A review of the current status of the Cabibbo-Kobayashi-Maskawa matrix (${CKM}$) is presented.
This paper is an update of the results published in \cite{ref:bello}. 
The experimental constraints imposed by the measurements of $\epsilonk$, $\vubovcb$, $\dmd$ and
from the limit on $\dms$ are used. Values of the constraints and of the parameters entering into the constraints, 
which restrict the range of the $\overline{\rho}$ and $\overline{\eta}$ parameters, include recent measurements presented at 
1998 Summer Conferences and progress obtained by lattice QCD collaborations.
The results are:
$$
\rhobar=0.202 ^{+0.053}_{-0.059}   ,~\etabar=0.340 \pm 0.035 
$$
from which the angles $\alpha$, $\beta$ and $\gamma$ of the unitarity triangle are inferred :
$$
\sin 2 \alpha =  -0.26 ^{+ 0.29}_{-0.28}  ,~\sin 2 \beta =  0.725 ^{+0.050}_{-0.060}   ,~\gamma= (59.5^{+8.5}_{-7.5})^{\circ} 
$$
Without using the constraint from $\epsilonk$, $\sin 2 \beta$ has been obtained: 
$\sin 2 \beta =   0.72 ^{+0.07}_{-0.11}$
Several external measurements or theoretical inputs have been removed, in turn, from the 
constraints and their respective probability density functions have been obtained.
Central values and uncertainties on these quantities  have been compared with actual measurements
or theoretical evaluations. In this way it is possible to quantify the importance of the different
measurements and the coherence of the Standard Model scenario for CP violation.
An important result is that $\Delta m_s$ is expected to be between [12.0 - 17.6]$ps^{-1}$ with 68$\%$ C.L. and 
$<20~ps^{-1}$ at 95$\%$ C.L. \\
Finally relations between the ${CKM}$ parameters and the quark masses are examined within a given model.
\end{abstract}

\section {Introduction.}
In a previous publication \cite{ref:bello}, uncertainties on the determination of the C.K.M.
parameters $A$, $\rhobar$ and $\etabar$, \footnote{$\rhobar$ and $\etabar$ are related to the original 
$\rho$ and $\eta$ parameters: $\rhobar(\etabar)= \rho(\eta)(1-\frac{\lambda^2}{2})$ 
\cite{ref:burbur}} in the framework of the Wolfenstein parametrization, have been reviewed.
This study was based on measurements and on theoretical estimates available at the beginning of 1997. Present results have been 
obtained using new measurements, coming mainly from LEP experiments and theoretical analyses available by end 1998. \\
Details on the determination of the values of the different 
quantities entering into the present analysis are explained in Section \ref{sec:2}.
Section \ref{sec:A} explains the evaluation of $A$ through the measurement of the $\Vcb$ element
of the C.K.M. matrix and, in Section \ref{sec:vub}, the new results on $\vubovcb$ from LEP and CLEO collaborations are presented.  
In Section \ref{sec:dms} the new limit on $\dms$ obtained by LEP experiments is recalled. Section \ref{sec:lattice} is 
dedicated to a review of the present determination of the non-perturbative QCD parameters, contributing to this analysis, 
from lattice QCD calculations. 
The present value of $f_{B_d}$, through the measurement of $f_{D_s}$, and the use of lattice QCD is explained \cite{ref:bello}. New 
results from lattice QCD relating the $\Bd$ and $\Bs$ decay constants are reported and finally the value used for $B_K$ is
commented.\\
Using the constraints on $\rhobar$ and $\etabar$ provided by the measurements
of $\epsilonk$, $\dmd$, $\dms$ and $\vubovcb$, in the framework of the Standard Model,
the region selected in the $(\rhobar,~\etabar)$ plane and the determination of the 
angles $\alpha$, $\beta$ and $\gamma$ of the unitarity triangle, are analyzed in Section \ref{sec:mesures}. 
Similar studies prior to the present analysis can be found in \cite{ref:alphabeta_old}. \\
Results have been also obtained by removing the constraint from $\epsilonk$ \cite{ref:barbieri}.
The information coming from one constraint or from an external parameters ($\dms$, $m_t$, $\Vcb$, $\vubovcb$, $B_K$, 
and $B_{B_d} \sqrt{f_{B_d}}$) has been removed, in turn, and the probability density distribution,
determined by the other parameters and constraints, has been determined in Section \ref{sec:param}.\\
Finally a test on a possible relation between the values of the ${CKM}$ matrix parameters and quark masses is shown in 
Section \ref{sec:masse}, using the framework of the parametrization of the C.K.M. matrix proposed in \cite{ref:friche}.

\section{Evaluation of the values of the parameters entering into this analysis.}
\label{sec:2}
In the following, two sets of values have been used for some of the parameters.
The first set corresponds to our best estimate (Scenario I). The purpose of using 
a second set (Scenario II) is to illustrate the variation of the uncertainties on the 
parameters with a more conservative evaluation of theoretical errors.

\subsection {The $A$ parameter.}
\label{sec:A}

The value of the parameter $A$ is obtained from the determination
 of $\Vcb$ in exclusive and inclusive semileptonic decays of
B hadrons. By definition:
\begin{equation}
\Vcb~=~A~ \lambda^2,~{\rm with}~\lambda~=~\sin \theta_c.
\end{equation}

Using exclusive decays $\Bbar \rightarrow \Dstar \ell^- \nubar$, the value
of 
$\Vcb$ is obtained by measuring the differential decay rate $\frac{d \Gamma_{D^{\ast}}}{d q^2}$
at maximum value of $q^2$ \cite{ref:neubertvcb1}.
$q^2$ is the mass of the charged lepton-neutrino system. At 
$q^2~=~q^2_{max}$, the $\Dstar$ is produced at rest in the B rest frame and HQET can be invoked
to obtain the value of the corresponding form factor: $F_{D^{\ast}}(w=1)$. The variable $w$ is usually
introduced; it is the product of the four-velocities of the B and $\Dstar$ mesons:
\begin{equation} 
w~=~v_B \cdot v_{D^\ast}~=~\frac{m_B^2+m_{D^\ast}^2-q^2}{2 m_B m_{D^\ast}},~w=1~{\rm for}~q^2=q^2_{max}.
\label{eq:2.01}
\end{equation}
In terms of $w$, the differential decay rate can be written:
\begin{equation} 
\frac{d BR_{D^{\ast}}}{d w}~=~\tau_{B^0_d} \frac{G_F^2}{48 \pi^3}m^3_{D^{\ast}}(m_B-m_{D^{\ast}})^2
K(w) \sqrt{w^2-1} F_{D^{\ast}}^2(w) \Vcb^2
\label{eq:2.02}
\end{equation}
in which $K(w)$ is a kinematic factor.

As the decay rate is zero for $w=1$, the $w$ dependence has to be adjusted over the measured range, 
using the previous expression and a parametrization of $F_{D^{\ast}}(w)$\cite{ref:neuneu}.

The following LEP average has been obtained:
\begin{equation} 
(F_{D^*}(1) \Vcb)~\times 10^3~=~ 35.3 \pm 1.7 \pm 1.7
\label{eq:2.03}
\end{equation}
which includes a new measurement from DELPHI \cite{ref:delphi2vcb}  based on a large sample of events and an improved
treatment of the contribution from $\Dstarstar$ decays. 
The accuracy of this measurement is still expected to improve by  adding new analyses from ALEPH and OPAL collaborations. 
In Table \ref{tab:vcbtab} the values measured by the different experiments included in this average have been reported. 
Central values have been corrected so that they correspond to the values of the parameters given in Table \ref{tab:constants}.
The statistical uncertainty on the average value of each $F_{D^*}(1) \times \Vcb$ measurement has been multiplied by 1.7 so 
that the corresponding $\chi^2/n.d.f.$ be equal to unity.

\begin{table}[htb]
\begin{center}
  \begin{tabular}{|c|c|c|c|}
    \hline
 Collaboration  &  Measured value (10$^3$)        &    Corrected value (10$^3$)    &  Ref.                  \\    \hline
   ALEPH        &     31.9 $\pm$ 1.8 $\pm$ 1.9    &    31.5 $\pm$ 1.8 $\pm$ 1.9    &   \cite{ref:alephvcb}  \\
   DELPHI1      &     35.4 $\pm$ 1.9 $\pm$ 2.4    &    35.0 $\pm$ 1.9 $\pm$ 2.4    &   \cite{ref:delphi1vcb}  \\
   DELPHI2      &     37.95 $\pm$ 1.34 $\pm$ 1.67 &    37.95 $\pm$ 1.34 $\pm$ 1.67 &   \cite{ref:delphi2vcb}  \\
   OPAL         &     32.8 $\pm$ 1.9 $\pm$ 2.2    &    32.4 $\pm$ 1.9 $\pm$ 2.2    &   \cite{ref:opalvcb}  \\
    \hline
  \end{tabular}
  \caption[]{\it { Measured and corrected values of the quantity $F_{D^*}(1) \times \Vcb$ included in the LEP average 
  (\ref{eq:2.03}). The corrected values have been obtained using the parameters given in Table \ref{tab:constants}. }}
  \label{tab:vcbtab}
\end{center}
\end{table}

Results have been combined, including the CLEOII measurement \cite{ref:cleovcb} and taking into account common 
systematics between LEP and CLEO results, coming from  $BR(\Do \rightarrow \Km \pi^+)$, $\tau_{B^0_d}$, 
$BR(\Dstarp \rightarrow \Do \pi^+)$ and the $\Dstarstar$ rate in semileptonic decays (see Tables \ref{tab:vcbtab}, 
\ref{tab:constants}). Using $F_{D^*}(1)=0.91 \pm 0.03$  \cite{ref:neubertvcb1}, $\Vcb$ is obtained from these exclusive 
measurements :

\begin{equation} 
( \Vcb)(exclusive)~\times 10^3~=~ 38.8 \pm 2.1 \pm 1.3 (theory)
\label{eq:2.04}
\end{equation}

Inclusive semileptonic decays of B hadrons can be used also to measure $\Vcb$
\cite{ref:ural} in the framework of the H.Q.E. ( Heavy Quark Expansion ) :

\begin{equation} 
( \Vcb)(inclusive)~\times 10^3~=~ 41.0 \pm 1.1 \pm 0.9 (theory)
\label{eq:2.05}
\end{equation}

The last error has been obtained by summing in quadrature the different theoretical contributions.
If these contributions are summed linearly the theoretical error has to be increased by a factor 1.5.
These two values of $\Vcb$ are in agreement and the corresponding theoretical uncertainties are uncorrelated.
The global average which has been used in the present analysis is:
\begin{equation} 
( \Vcb)~\times 10^3~=~ 40.4 \pm 1.2
\label{eq:2.06}
\end{equation}
which gives :
\begin{equation}
 A = 0.831  \pm  0.029
\label{eq:Aopt}
\end{equation}
It has to be noticed that this result does not yet constitute really an accurate measurement because, experimentally, the present 
uncertainty corresponds to a 6$\%$ error on measured quantities. It has to be also stressed that there is not yet a general 
consensus, among theorists, concerning the importance of theoretical errors affecting $\Vcb$. As experimental result become more 
and more accurate this point needs to be clarified soon. 

The more conservative evaluation of the uncertainty on $\Vcb$,
has been obtained by multiplying theoretical errors by 1.5 :

\begin{equation}
( \Vcb)~\times 10^3~=~ 40.4 \pm 1.5
\label{eq:2.06bis}
\end{equation}

which gives :

\begin{equation}
 A = 0.831  \pm  0.035
\label{eq:Aoptbis}
\end{equation}

This last evaluation will be used in the second scenario.
In \cite{ref:pdg98} the following value is given :

\begin{equation}
( \Vcb)~\times 10^3~=~ 39.5 \pm 1.7
\label{eq:2.06tris}
\end{equation}

It agrees with the present evaluation and the quoted error is slightly larger than the present values
(\ref{eq:2.06},\ref{eq:2.06bis}) because this average does not include more recent measurements.

\begin{table}[htb]
\begin{center}
  \begin{tabular}{|c|c|c|}
    \hline
 Parameter &  Value     &    Ref.  \\
    \hline
 ${\rm BR}(\Do \rightarrow \Km \pi^+)$ & $ (3.85 \pm 0.09)\%$     & \cite{ref:pdg98}     \\
 $\tau_{B^0_d}$ & $ (1.57  \pm 0.04) ps $     & \cite{ref:taugroup}   \\
 ${\rm BR}(\Bd \rightarrow \ell X)$    & $(10.2 \pm 0.5)\%$ & (see legenda) \\
${\rm P}(b \rightarrow \Bd)$ & $ (39.5 ^{+1.6}_{-2.0})\%$ & \cite{ref:osciw}     \\
${\rm R}_b$ & $ 21.58 \%$ & Std. Model \\
    \hline
  \end{tabular}
  \caption[]{\it {Values of the parameters used in the present determination of $\Vcb$.
The semileptonic branching fraction for the $\Bd$ meson has been obtained using the inclusive
semileptonic branching fraction measurement done at the $\Upsilon$(4S) \cite{ref:brcleo} and correcting for the
contribution of charged B mesons by taking into account the difference between $\Bd$
and $\Bm$ lifetimes.}}
  \label{tab:constants}
\end{center}
\end{table}

\subsection{Present value of $\Vub$/$\Vcb$.}
\label{sec:vub}

LEP Collaborations \cite{ref:vublep} have provided new results on $\Vub$
which have different systematics and a similar accuracy as the inclusive CLEO measurement ( using the endpoint of the 
lepton spectrum (inclusive) \cite{ref:cleoincl}).  Using the value of $ \Vcb$ given in (\ref{eq:2.06}) it follows :
\begin{equation} 
\frac{\Vub}{\Vcb}~=~ 0.104 \pm 0.011 (exp.) \pm 0.015 (theo.) ~~~~~~ \rm{LEP ~~average}
\end{equation}
A recent result has been obtained by the CLEO Collaboration on $\Vub$ by measuring the $B \rightarrow \pi \ell \nu$ and 
$B \rightarrow \rho(\omega) \ell \nu$ branching fractions \cite{ref:cleoexcl}. Using the value of $ \Vcb$ given in (\ref{eq:2.06})
it follows :
\begin{equation} 
\frac{\Vub}{\Vcb}~=~ ( 0.081 \pm 0.011 \pm 0.029 (theo.) ) ~~~\rm{CLEO-exclusive ~~~B \rightarrow \pi(\rho,\omega) \ell \nu}
\end{equation}
The quoted theoretical error is estimated to be the maximal excursion of the values of $\frac{\Vub}{\Vcb}$ 
evaluated in different models \cite{ref:vubmodels}. From the measured ratio of 
$\frac{\rho}{\pi}$ production ratio the (KS) model is disfavoured \cite{ref:stone}. 
This is quite important since the value of $\Vub$ obtained in this model deviated 
the most from the other estimates both in the exclusive and inclusive measurements \cite{ref:cleoincl}. 
After having removed this model, the maximal excursion on the values of $\frac{\Vub}{\Vcb}$, evaluated in different 
models from the endpoint measurement (CLEO-inclusive), is $ 0.008 $. 
A conservative approach has been used in this paper which consists in doubling this error. The result is then :
\begin{equation} 
\frac{\Vub}{\Vcb}~=~ 0.080 \pm 0.006 \pm 0.016 (theo.) ~~~~~~ \rm{CLEO - inclusive}
\end{equation}
In the following, only the results from LEP and from the inclusive CLEO measurement are used. Theoretical errors between 
these two measurements are largely uncorrelated and the two results will be used as independent constraints on the ratio 
$\frac{\Vub}{\Vcb}$.

\subsection{Present limit on {$\Delta m_s$}.}
\label{sec:dms}

A new limit on $\Delta m_s$, $\Delta m_s > 12.4 ps^{-1} ~{\rm at}~95 \% C.L.$,
has been derived by the "B Oscillation Working Group" \cite{ref:osciw}. The sensitivity 
of present measurements is at $13.8 ps^{-1}$. \\
This limit has been obtained in the framework of the amplitude method \cite{ref:amplitude} which
consists in measuring, for each value of the frequency $\Delta m_s$, an amplitude $a$ and its error $\sigma(a)$.
The parameter $a$ is introduced in the time evolution of pure $\mbox{B}^0_s$ or $\overline{\mbox{B}^0_s}$ states so that the
value $a=1$ corresponds to a genuine signal for oscillation:
$$
{\cal P}(\mbox{B}^0_s\rightarrow (\mbox{B}^0_s,~\overline{\mbox{B}^0_s}))~
=~\frac{1}{2 \tau_{s}} e^{- \frac{t}{\tau_{s}}} \times
 ( 1 \pm a~cos ({\Delta m_s t} ) )
$$
The values of $\Delta \rm{m}_s$ excluded at 95\% C.L. are those satisfying the condition a($\Delta \rm{m}_s$) + 1.645 $\sigma_a
(\Delta \rm{m}_s) < 1$. With this method it is easy to combine different experiments and to treat systematic uncertainties 
in an usual way since, at each value of $\Delta \rm{m}_s$, a value for a with a Gaussian error $\sigma_a$, is measured. 
Furthermore the sensitivity of the experiment can be defined as the value of
$\Delta \rm{m}_s$ corresponding to 1.645 $\sigma_a (\Delta \rm{m}_s) = 1$ (for
a($\Delta \rm{m}_s) = 0$, namely supposing that the ``true'' value of $\Delta \rm{m}_s$ is well above the limit. The
sensitivity is the limit which would be reached in 50\% of the experiments. \\
However the set of measurements $a(\Delta m_s)$ contains more information than the $95 \%$ C.L. limit. It is possible
to build a $\chi^2$, which quantifies the compatibility of $a(\Delta m_s)$ with the value $a=1$, defined as:
$$
\chi^2(\Delta m_s)  = {(a(\Delta m_s)-1)^2 \over \sigma^2(a(\Delta m_s))^2}
$$
and which can be used as a constraint.\\

\subsection{ Present values for the non-perturbative QCD parameters.}
\label{sec:lattice}

Important improvements have been achieved in the last few years in the evaluation of the non-perturbative QCD parameters, entering
into this analysis, in the
framework of lattice QCD. As a consequence, in this paper, only most recent results are used. 
Motivations for this attitude have been given in \cite{ref:kronefeld1}.
Recent reviews on lattice QCD can be found in \cite{ref:kronefeld1},\cite{ref:review1},\cite{ref:review2}.

\subsubsection{Present value of {$f_{B_d}$}.}
\label{sec:fb}
\begin{figure}
\begin{center}
{\epsfig{figure=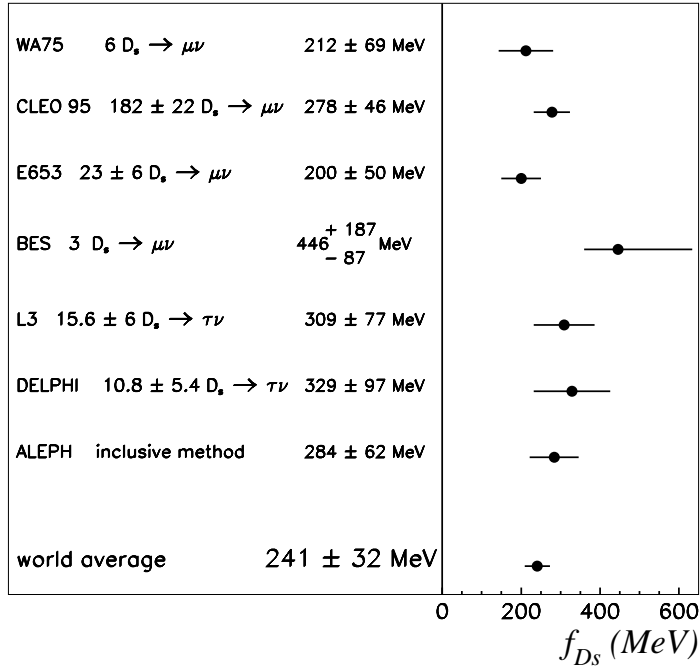,bbllx=0pt,bburx=620pt,bblly=0pt,bbury=521pt,height=8cm}}
\caption{ \it { Summary of the evaluation of $f_{D_s}$ from the measurements of $Br(D_s \rightarrow \mu \nu)$ and 
$Br(D_s \rightarrow \tau \nu)$. }}
\label{fig:fds}
\end{center}
\end{figure}

Following the proposal made in \cite{ref:bello}, $f_{B_d}$ is evaluated from the measurements of $f_{D_{s}}$ and 
using the extrapolation from the D to the B sector as predicted by lattice QCD \cite{ref:soni}. 
The value of $f_{D_{s}}$ is deduced from the measurements of the branching fractions : $\rm{D^+_s \rightarrow \tau^{+} \nu_{\tau}}$
 and $\rm{D^+_s \rightarrow \mu^{+} \nu_{\mu}}$. The different determinations of  $f_{D_{s}}$ \cite{ref:dstaunu} are shown in 
Figure~\ref{fig:fds}. The recent measurements from CLEO and ALEPH collaborations have been included \cite{ref:dstaunu}.

The average is:

\begin{equation}
f_{D_{s}} = ( 241 \pm 32 )~ MeV
\label{eq:fds}
\end{equation}

This value is in agreement with those most recently obtained from lattice QCD calculations, quoted in Table~\ref{tab:qcdparam}.

\begin{table}[htb]
\begin{tabular}{|c|c|c|c|c|}
\hline
$f_{D_s}$ [MeV]     & $f_{B_d}$ [MeV]    &    $f_{B_d} / f_{D_s} $  &     $f_{{B}_s}/f_{{B}_d}$  &   ref. \\ \hline
$224 \pm 2 \pm 16 \pm 12 $    & $173 \pm 9 \pm 8 \pm 11$  &   &   &   \cite{ref:aoki} \\
$213 \pm ^{+14}_{-11} \pm 11 ^{+21}_{-0} $ & $164 ^{+14}_{-11} \pm 8 ~^{+16}_{-0}$ &   $0.78 \pm 0.04$ & 1.13 $^{+5}_{-4}$ &  
\cite{ref:kronefeld} \\
$213 \pm 9 ^{+23}_{-9} {^{+17}_{-0}}    $    & $159 \pm 11 ^{+22}_{-9} {^{+21}_{-0}}$  & $0.75 \pm 0.03^{+0.04~+0.07}_{-0.02~-0.00}$ & 
1.10 $\pm 0.02 {^{+0.05}_{-0.03}} {^{+0.03}_{-0.02}} $  &  \cite{ref:milc}    \\
$231 \pm 12 ^{+6}_{-1} $ & $179 \pm 17 ^{+26}_{-9}$ & $0.71 \pm 0.04^{+0.07}_{-0.00}$ & 1.14 $\pm 0.03 ^{+0.00}_{-0.01}$ &  
\cite{ref:apeape}  \\
                   & $147 \pm 11 \pm 11 ^{+8}_{-12}$  &  &  &  \cite{ref:glok} \\   
\hline
\end{tabular}
\caption[]{ \it {  $f_{D_s}$, $f_{B_d}$, $f_{B_d} / f_{D_s}$ and $f_{{B}_s}/f_{{B}_d}$  parameters from recent lattice QCD calculations. 
For the $f_{D_s}$ and $f_{B_d}$ parameters the first two errors are of statistical and systematical origins respectively, the 
third comes from the use of the quenched approximation. }}
\label{tab:qcdparam}
\end{table}

In recent publications, from lattice QCD, the ratio between the $\Ds$ and the $\Bd$ decay constants has been 
evaluated, results are also given in Table~\ref{tab:qcdparam}. 
All evaluations are in agreement. The errors coming from 
the quenched approximation are estimated to be of the order of 10$\%$ \cite{ref:milc},\cite{ref:apeape} and tend to increase 
the value of the ratio $f_{B_d}/f_{D_s} $.

Using the experimental value $f_{D_{s}} = ( 241 \pm 32 )~ MeV$, and the theoretical evaluation given in 
Table~\ref{tab:qcdparam} from \cite{ref:milc}, it follows :

\begin{equation}
f_{B_d} = 181 \pm 24(exp.) \pm 7(theo. stat.) ^{+20}_{-5} (theo.~~non.~~stat.) MeV
\label{eq:fbded}
\end{equation}
This result can be compared with the most recent evaluations of $f_{B_d}$, from lattice QCD, which are also given in 
Table~\ref{tab:qcdparam}. The values obtained for $f_{B_d}$ are in relative agreement and 
are also in agreement with the extrapolation from the $f_{D_s}$ measurement (\ref{eq:fbded}).
In the following, the value used for $f_{B_d}$ is the one coming from the extrapolation of the $f_{D_s}$ measurement 
(\ref{eq:fbded}). \\

\subsubsection{ Present value of $f_{B_d} \sqrt {B_{B_d}}$. }

Present evaluations of the ${B_{B_d}}$ parameter are given in Table \ref{tab:bb}. With respect to the situation concerning
the previously discussed parameters, ${f_{D_s}}$ and ${f_{B_d}}$, most recent results on ${B_{B_d}}$ are not nicely in agreement. 
On the other hand the error coming from the quenched approximation evaluated in \cite{ref:soni} is small, of the order of 4$\%$. \\ 
A conservative approach consists in using :
\begin{equation}
    B_{B_d} = 1.35 \pm 0.15
\label{eq:bbresult}
\end{equation}

Combining this result with the value of ${f_{B_d}}$ given in (\ref{eq:fbded}) the value used in this analsyis for 
${f_{{B}_d} \sqrt{B_{B_d}}}$ is :

\begin{equation}
f_{B_d} \sqrt{B_{B_d}}~=~ 210 \pm 29 (stat.) ^{+23}_{-6} (theo.~~non~~stat) \pm 12 (from ~B_{B_d})  ~MeV
\label{eq:fbsqrtb}
\end{equation}

\begin{table}[htb]
\begin{center}
\begin{tabular}{|c|c|}
\hline
$ B_{B_d} $  &  Ref. \\
\hline
 $1.53 \pm 0.19$                         &  \cite{ref:bbs}       \\
 $1.42 \pm 0.07 $                        &  \cite{ref:aokibb}  \\
 $1.44 \pm 0.09 \pm 0.06$                &  \cite{ref:soni}    \\
 $1.29 \pm 0.08 \pm 0.06 $               &  \cite{ref:reyesgime} \\ 
 $1.40 \pm 0.06 ^{+0.04}_{-0.26}$        &  \cite{ref:cdm} \\ 
 $1.17 \pm 0.09  \pm 0.05  $             &  \cite{ref:yamada} \\ 
\hline
\end{tabular}
\caption[]{ \it { Values of the ${B_{B_d}}$ parameter from recent lattice QCD calculations. The second error in \cite{ref:soni}
is an estimate of the error which comes from the use of the quenched approximation. }}
\label{tab:bb}
\end{center}
\end{table}

In table \ref{tab:fbpre} the contributions of different parameters to the error on $ f_{B_d}\sqrt{B_{B_d}}$ are given.

\begin{table}[htb]
\begin{center}
\begin{tabular}{|c|c|}
\hline
 parameter (error)  &  error on $f_{B_d} \sqrt{B_{B_d}}$ \\
\hline
 $ f_{D_s} $ (stat.)  ($16\%$)                          &        $8\%$             \\
 $ Br(D_s \rightarrow \phi \pi) $  ($25\%$)             &        $10\%$            \\ 
  $ other~ systematics $                                &        $5\%$             \\
 $f_{B_d}/f_{D_s}$ (stat.)  ($4\%$)                     &        $4\%$             \\
 $f_{B_d}/f_{D_s}$(non stat.)  ($^{+11\%}_{-3\%}$)      &     $^{+11\%}_{-3\%}$    \\
      $ B_{B_d}  $  ($11\%$)                                &        $6\%$             \\
\hline
\end{tabular}
\caption[]{ \it {   Contributions to the precision on $f_{B_d} \sqrt{B_{B_d}}$ from different quantities. The main contribution in 
$"other~systematics"$ comes from the assumption that the mechanism of production of strange B and D mesons in jets is the same.}}
\label{tab:fbpre}
\end{center}
\end{table}

\subsubsection {Present value for $\xi$.}
\label{sec:fbsofbd}
Significant improvements have been achieved in the determination of the $\xi$ parameter, defined as 
$\frac{f_{B_s} \sqrt{B_{B_s}}}{f_{B_d} \sqrt{B_{B_d}}}$. Several authors agree
on a relative precision better than 10$\%$ on the ratio $f_{{B}_s}/f_{{B}_d}$ (see Table \ref{tab:qcdparam}). 
The error coming from the quenched approximation 
seems to be controlled at the level of 3$\%$. The ratio of the bag factors $\frac{B_{B_s}}{B_{B_d}}$ is known very precisely 
( $\frac{B_{B_s}}{B_{B_d}} = 1.01 \pm 0.01$ \cite{ref:gimemarti} ).

The value from reference \cite{ref:milc} has been used in the following :
\begin{equation}
\xi = 1.11 \pm 0.02 ^{+0.06}_{-0.04}
\label{eq:xi}
\end{equation}

\subsubsection{Present value of $B_K$.}


The two most recent values are in agreement :
\begin{eqnarray}
B_K( ~\rm{at}~ 2 ~ GeV ) &=& 0.62 \pm 0.02 \pm 0.02   \cite{ref:gupta} \nonumber \\
                         &=& 0.628 \pm 0.042           \cite{ref:aokibk} 
\label{resultbk}
\end{eqnarray}

The scale-invariant value for $B_K$ is then evaluated to be \cite{ref:gupta}: $B_K = 0.86 \pm 0.03 \pm 0.03$. 
The error due to the quenched approximation is evaluated to be less than $10\%$. For this analysis the following value is used :

\begin{equation}
B_K = 0.86 \pm 0.06 \pm 0.08 ( theo.~~ from~~ quenching ) 
\label{eq:bk1}
\end{equation} 

A more conservative evaluation of theoretical uncertainties has been proposed in \cite{ref:review2}. Differences
on the central value and on the last error with respect to (\ref{eq:bk1}) are coming from the evaluation of the 
quenched approximation. The value is:
\begin{equation}
B_K = 0.94 \pm 0.06 \pm 0.14 ( theo.~~ from~~ quenching )
\label{eq:bk2}
\end{equation}
This second evaluation has been used in the conservative scenario (Scenario II).

\begin{table}[htb]
\begin{center}
\frame{\footnotesize {
\begin{tabular}{|c|c|c|c|}
\hline
                       parameter                 &  Values                      & Gaussian $-$ Flat error   &  Ref. \\ 
\hline 
                    $\lambda$                    & $0.2205 \pm 0.0018$          & $\pm 0.0018 - \pm 0.000$  & \cite{ref:bello}  \\
$\left | V_{cb} \right |$(Scenario I)       & $(40.4 \pm 1.2)\times 10^{-3}$   & $(\pm 1.2 - \pm 0.0)\times 10^{-3}$ & this paper        \\
$\left | V_{cb} \right |$(Scenario II)     & $(40.5 \pm 1.5)\times 10^{-3}$   & $(\pm 1.5 - \pm 0.0)\times 10^{-3}$ & this paper        \\
$\frac{\left | V_{ub} \right |}{\left | V_{cb} \right |}$ (CLEO) & $0.080 \pm 0.017$ & $\pm 0.006 - \pm 0.016$ & this paper     \\
$\frac{\left | V_{ub} \right |}{\left | V_{cb} \right |}$ (LEP)  & $0.104 \pm 0.019$ & $\pm 0.011 - \pm 0.015$ & this paper     \\
                    $\Delta m_d$                 & $(0.472 \pm 0.016) ~ps^{-1}$ & $\pm 0.016 - \pm 0.000$   & \cite{ref:osciw}  \\
                    $\Delta m_s$                 & $>$ 12.4 ps$^{-1}~{\rm at}~95 \% ~C.L.$ & see text (\ref{sec:dms})  & \cite{ref:osciw}  \\
               $ \overline{m_t}(m_t)$            & $(167 \pm 5) ~GeV/c^2$       & $ \pm 5  - \pm 0$         & \cite{ref:topmass} \\
               $B_K$ (Scenario I)                & $0.86 \pm 0.09$               & $\pm 0.06  -  \pm 0.08$  & this paper     \\
               $B_K$ (Scenario II)               & $0.94 \pm 0.15$               & $\pm 0.06  -  \pm 0.14$  & this paper     \\
               $ f_{B_d} \sqrt{B_{B_d}}$      & $(210^{+39}_{-32})~MeV$ & $\pm 29  - ^{+23}_{-6}~\rm{and}~ \pm 12$  & this paper  \\
$\xi=\frac{ f_{B_s}\sqrt{B_{B_s}}}{ f_{B_d}\sqrt{B_{B_d}}}$ & $1.11^{+0.06}_{-0.04}$ &$\pm 0.02 - ^{+0.06}_{-0.04}$ & this paper  \\
$\overline{m_c}(m_c)$                         & $(1.3 \pm 0.1)~GeV/c^2 $      & $\pm 0.1 - \pm 0.000$  & \cite{ref:pdg98} \\
$\eta_1$                                      &  $1.38 \pm 0.53$              & $\pm 0.53 - \pm 0.000$ & \cite{ref:buras1} \\
$\eta_3$                                      &  $0.47 \pm 0.04$              & $\pm 0.04 - \pm 0.000$ & \cite{ref:buras1} \\
$\eta_2$                                      &  $0.574 \pm 0.004$            &        fixed           & \cite{ref:buras1} \\
$f_K$                                         & $0.161~GeV/c^2$               &        fixed           & \cite{ref:pdg98} \\
$\Delta m_K$                                  & $(0.5301 \pm 0.0014) \times 10^{-2} ~ps^{-1}$  & fixed   & \cite{ref:pdg98} \\
$\epsilonk$                        & $(2.280\pm 0.019) \times 10^{-3}$ &   fixed           & \cite{ref:pdg98} \\
$\eta_B$                                      & $0.55 \pm 0.01$               &        fixed           & \cite{ref:buras1} \\
$ G_F $                                       & $(1.16639 \pm 0.00001)\times 10^{-5} GeV^{-2}$  & fixed &\cite{ref:pdg98} \\
$ m_{W}$                                      & $80.41 \pm 0.10 ~GeV/c^2 $     &       fixed           &  \cite{ref:pdg98} \\
$ m_{B^{0_d}}$                                & $5.2792 \pm 0.0018 ~GeV/c^2 $  &        fixed          & \cite{ref:pdg98} \\
$ m_{B^{0_s}}$                                & $5.3693 \pm 0.0020 ~GeV/c^2 $  &       fixed           &  \cite{ref:pdg98} \\
$ m_K$                                        & $0.493677 \pm 0.000016 ~GeV/c^2$ &        fixed          & \cite{ref:pdg98} \\ 
\hline 
\end{tabular} }}
\caption[]{ \it {Values of the quantities entering into the expressions of $\epsilonk$, $\vubovcb$, $\dmd$ and $\dms$.
In the third column the Gaussian and the flat part of the error are given explicitely. Two scenarios are considered depending 
on the expected theoretical uncertainties on the values of the parameters $B_K$ and $\Vcb$. }} 
\label{tab:a} 
\end{center}
\end{table} 

\section{Results with present measurements. }
\label{sec:mesures}

The central values and uncertainties for the relevant parameters used in this analysis are given in Table \ref{tab:a}. \\
For the ${\left | V_{cb} \right |}$ and $B_K$ parameters two sets of values are quoted corresponding to the present best estimate 
(scenario I) and to a more conservative evaluation of theoretical uncertainties ( scenario II). In this paper all
figures correspond to scenario I, but numerical results are given for both scenarios. \\
It has to be reminded that the approach consists in building up the two dimensional probability distribution for
$\overline{\rho}$ and $\overline{\eta}$ \cite{ref:bello}. This is done as follows :
\begin{itemize}
\item
a point uniformly distributed in the ($\overline{\rho},\overline{\eta}$) plane, is chosen,
\item
values for the different parameters entering into the equations of constraints
are obtained using random generations extracted from Gaussian/uniform distributions depending on the source of the error. 
As an example, for the non-perturbative QCD parameters ( as ${f_{B_d} \sqrt{B_{B_d}}}$, $\xi$, ${B_K}$ ) the error coming 
from the quenching approximation is extracted from a flat distribution.
\item
the predicted values for the four quantities,
$ \frac{\left | V_{ub} \right |}{\left | V_{cb} \right |}$, $\epsilonk$\footnote{the formula given in \cite{ref:bello} 
has been corrected to $\epsilonk~ =~ C_{\epsilon}~ \mbox{B}_K~A^2 \lambda^6 \overline{\eta} [-\eta _1 
(1-\frac{\lambda^2}{2})S(x_c)+A^2 \lambda^4(1-\overline{\rho}-(\overline{\rho}^2+\overline{\eta}^2-\overline{\rho})\lambda^2) 
\eta _2 S(x_t) + \eta _3 S(x_c,x_t)].$ The term $\overline{\rho}$ was missing in the
$(\overline{\rho}^2+\overline{\eta}^2-\overline{\rho})$ term.}, $\Delta m_d$ and
$\frac{\Delta m_d}{\Delta m_s}$ are then obtained and compared with present measurements which can have Gaussian/non Gaussian 
errors. As an example, for the constraint provided by the measurement of $ \frac{\left | V_{ub} \right |}{\left | V_{cb} \right|}$,
theoretical errors are treated as a flat distribution. 
A weight is then computed which takes into account the expected shape of the probability distribution for the constraint 
(Gaussian, flat or a convolution of the two distributions). The final weight is equal to the product of all independent weights.
\item
the sum of all weights, over the ($\overline{\rho},\overline{\eta}$) plane, is normalized to unity and contours
corresponding to 68$\%$ and 95$\%$ confidence levels have been defined.
\end{itemize}

\begin{figure}
\begin{center}
{\epsfig{figure=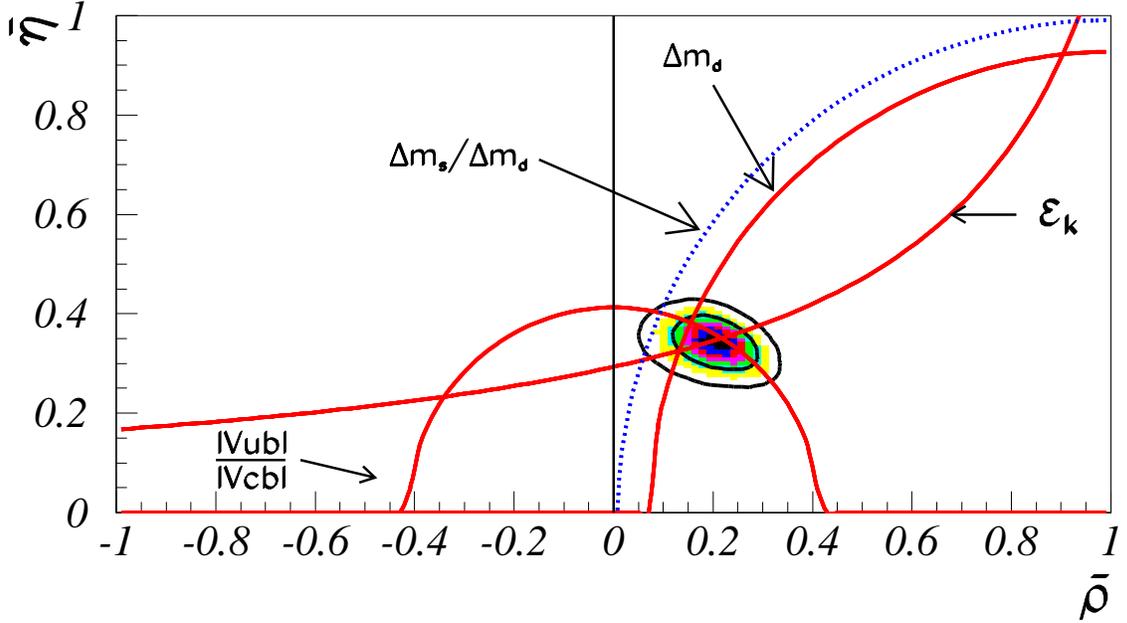,bbllx=30pt,bburx=503pt,bblly=8pt,bbury=249pt,height=8cm}}
\caption{ \it{ The allowed region for $\overline{\rho}$ and $\overline{\eta}$ using the parameters listed in Table \ref{tab:a}.
The contours at 68 $\%$ and 95 $\%$ are shown. The full lines correspond to the central values of the constraints given by
the measurements of  $  \frac{\left | V_{ub} \right |}{\left | V_{cb} \right |} $, $\epsilonk$ and $\Delta m_d$.
The dotted curve corresponds to the 95 $\%$ C.L. upper limit obtained from the experimental limit on
$\Delta m_s$.  }}
\label{fig:rhoeta_cimento}
\end{center}
\end{figure}
 
The region of the $(\rhobar,~\etabar)$ plane selected by the measurements of $\epsilonk$,
$\vubovcb$, $\dmd$ and from the limit on $\dms$ has been obtained and is given in Figure \ref{fig:rhoeta_cimento}.
The measured values for the two parameters are:

\begin{equation}
\rhobar=0.202 ^{+0.053}_{-0.059}   ,~\etabar=0.340 \pm 0.035 ~~~~~~~~~~~~ \rm{Scenario~~ I} 
\label{eq:rhoeta1}
\end{equation}

\begin{equation}
\rhobar=0.214 ^{+0.055}_{-0.062}   ,~\etabar=0.328 \pm 0.040 ~~~~~~~~~~~~ \rm{Scenario~~ II}
\label{eq:rhoeta2}
\end{equation}

The errors on $\rhobar$ and $\etabar$, between the two scenarios, differ by about 10$\%$.\\
It is clear that the allowed region for $\rhobar$ is not symmetric around zero, negative values for $\rhobar$ being clearly
disfavoured : ${\cal P}_{\rho<0}=  0.8 \%$. 

\subsection{Measured values of $\rhobar$ and $\etabar$ without the $\epsilonk$ constraint.}
\label{sec:senzaepsk}

\begin{figure}
\begin{center} 
{\epsfig{figure=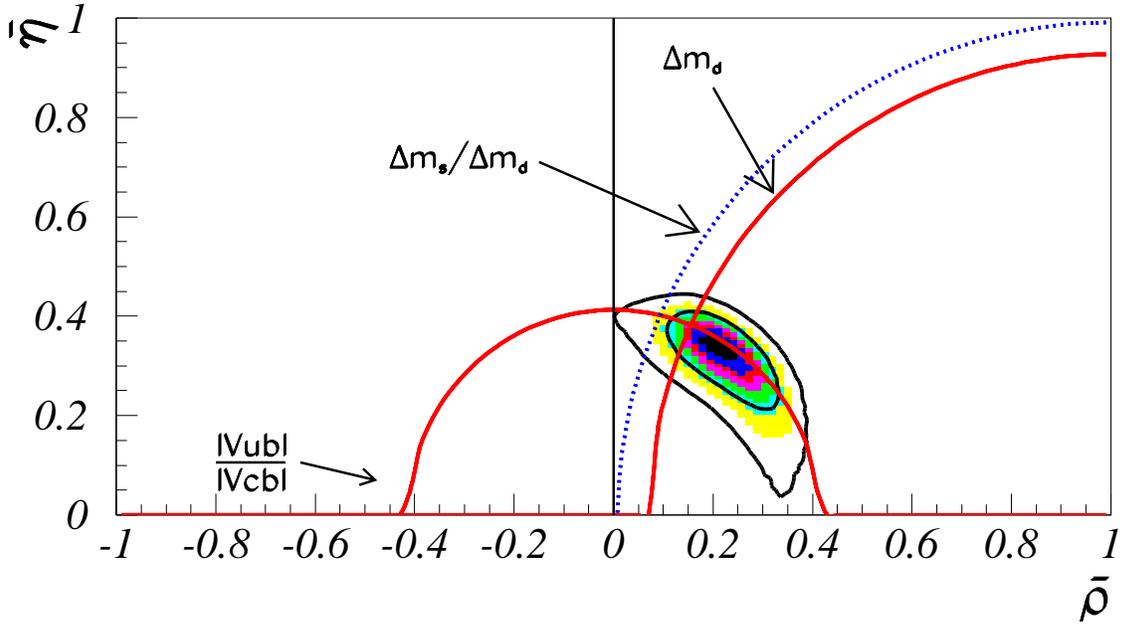,bbllx=30pt,bburx=503pt,bblly=8pt,bbury=249pt,
height=8cm}}
\caption{ \it{ The allowed region for $\overline{\rho}$ and $\overline{\eta}$ using the parameters listed in Table \ref{tab:a}
without using the $\epsilonk$ constraint. The contours at 68 $\%$ and 95 $\%$ are shown. The dotted curve 
corresponds to the 95 $\%$ C.L. upper limit obtained from the experimental limit on $\Delta m_s$.  }}
\label{fig:rhoeta_senzaepsk}
\end{center}
\end{figure}

Following the idea proposed in \cite{ref:barbieri}, a region of the $(\rhobar,~\etabar)$ plane can be selected without using the 
$\epsilonk$  constraint. The result is shown in Figure~\ref{fig:rhoeta_senzaepsk}, where the contours corresponding to 
68 $\%$ and 95 $\%$ confidence levels are also indicated. This test shows that the $(\rhobar,
~\etabar)$  region selected by the measurements in the B sector is well compatible with the region selected from the
measurement of the direct CP violation in the kaon sector. \\
The value $\etabar = 0 $ is situated in the region excluded at 96$\%$ C.L. in scenario I and at 95$\%$ C.L. in scenario II.
Another approach which consists in testing the hypothesis of a real C.K.M matrix by setting $\etabar$=0 can be found in 
\cite{ref:checchia}. 

\subsection{ Measured values of $\sin 2\alpha$, $\sin 2\beta$ and $\gamma$.}

It is of interest to determine the central values and the uncertainties on the quantities
$\sin 2\alpha$, $\sin 2\beta$ and $\gamma$ which can be measured directly at future facilities like, CDF, D0,
HERA-B, B factories and LHC experiments. The results are :

\begin{figure}
\begin{center}
{\epsfig{figure=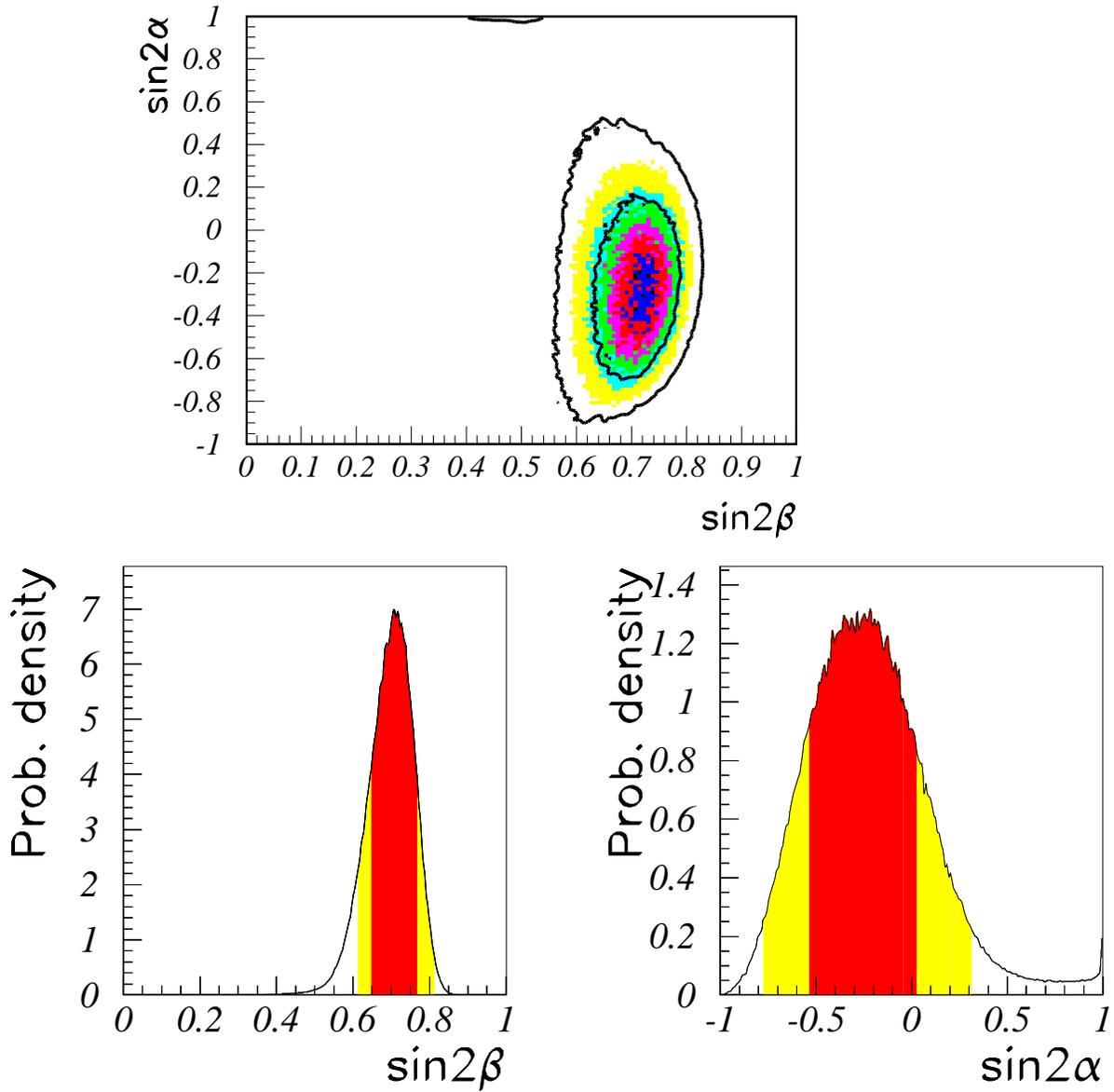,bbllx=0pt,bburx=620pt,bblly=0pt,bbury=521pt,height=16cm}}
\caption{ \it{ The $sin 2 \alpha$ and $sin 2 \beta$ probability distributions have been obtained using the contraints corresponding 
to the values of the parameters listed in Table \ref{tab:a}.  The dark-shaded and the clear-shaded intervals correspond, 
respectively, to 68$\%$ and 95 $\%$ confidence level regions. }}
\label{fig:seni_cimento}
\end{center}
\end{figure}

\begin{equation}
\sin 2 \alpha =  -0.26 ^{+ 0.29}_{-0.28}  ,~\sin 2 \beta =  0.725 ^{+0.050}_{-0.060}   ,~\gamma= (59.5^{+8.5}_{-7.5})^{\circ} 
 ~~~~~~~~~~~~ \rm{Scenario~~ I} 
\label{eq:abc1}
\end{equation}

\begin{equation}
\sin 2 \alpha =  -0.36 ^{+ 0.33}_{-0.30}  ,~\sin 2 \beta = 0.715 ^{+0.055}_{-0.065}  ,~\gamma= (56.5^{+9.5}_{-8.5})^{\circ} 
~~~~~~~~~~~~ \rm{Scenario~~ II} 
\label{eq:abc2}
\end{equation}

Figure \ref{fig:seni_cimento} gives the correlation between the measurements of $\sin 2 \alpha$ and $\sin 2 \beta $ and the 
contours at 68$\%$ and 95 $\%$ C.L.. Figure \ref{fig:haricot2_gamma} gives the distribution of the angle $\gamma$.\\
The preliminary result from the CDF Collaboration is $\sin 2 \beta =  0.79 ^{+0.41}_{-0.44}$ \cite{ref:cdfsin2beta}.

\subsubsection{Measurement of $sin 2 \beta$.}
The value of $sin 2 \beta$ is rather precisely determined, with an accuracy already at a level expected after 3-4 years of
running at B factories. The situation will improve in the current year (1999) with better
measurements of $\Vcb$, with a possible improvement of the sensitivity of LEP analyses on $\Delta m_s$ and with 
an expected progress from lattice QCD calculations. \\
Without using the constraint on $\epsilonk$ : $\sin 2 \beta =  0.72 ^{+0.07}_{-0.11}$.

\subsubsection{Measurement of $sin 2 \alpha$.}

In our previous analysis \cite{ref:bello}, it was concluded that there was no restriction on the domain
of variation of $sin 2 \alpha$ between -1 and +1. The present study, see Figure \ref{fig:seni_cimento},
allows to identify now a favoured domain for this parameter.

\subsubsection{Measurement of the angle $\gamma$.}
 It has been proposed in \cite{ref:flman} to restrict the range of variation of the angle $\gamma$
using the measurement of the ratio, $R_1$, of the branching fractions of charged and neutral B mesons
into ${\rm K} \pi$ final states. In the hypothesis that this ratio is below unity, the following
constraint has to be satisfied:
\begin{equation}
\sin ^2 \gamma < R_1
\end{equation}
 
\begin{figure}
\begin{center}
{\epsfig{figure=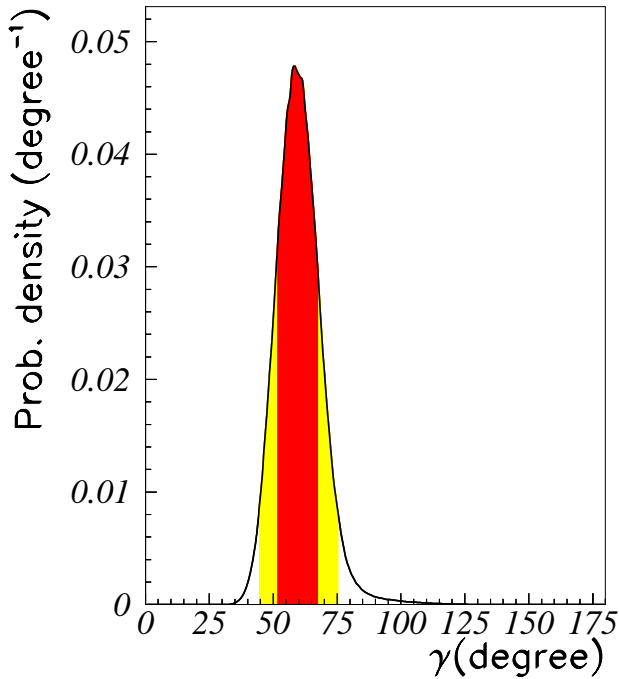,bbllx=0pt,bburx=470pt,bblly=0pt,bbury=515pt,height=9cm}}
\caption{ \it{ The $\gamma$ angle probability distribution obtained using the same constraints as in Figure 
\ref{fig:rhoeta_cimento}. The dark-shaded and the clear-shaded intervals correspond to 68$\%$ and 95 $\%$ confidence level 
regions respectively.  }}
\label{fig:haricot2_gamma}
\end{center}
\end{figure}

\noindent
The present result from CLEO \cite{ref:cleokpi}:
\begin{equation}
R_1~=~\fleisher~=0.65 \pm 0.40,
\end{equation}
has a too large uncertainty to be really constraining on $\gamma$. This bound excludes a region
which is symmetric around $\gamma = 90^{\circ}$. 
In fact, as explained already in Section \ref{sec:mesures},
negative values of $\rhobar$ are already excluded and the region around $\gamma = 90^{\circ}$ has
a low probability. These restrictions are clearly apparent in Figure \ref{fig:haricot2_gamma} which 
gives the expected density probability distribution for the angle $\gamma$, which is determined to 
be : $\gamma = (59.5^{+8.5}_{-7.5})^{\circ}$

\begin{figure}
\begin{center}
{\epsfig{figure=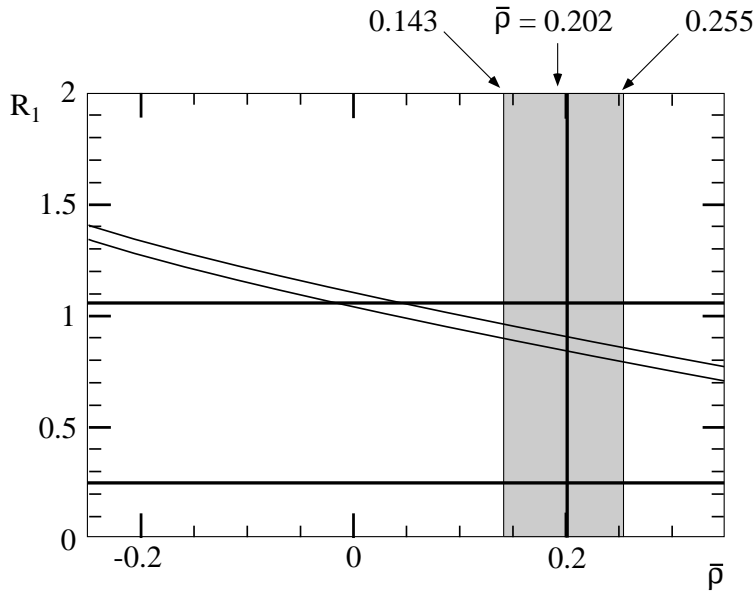,bbllx=76pt,bburx=413pt,bblly=80pt,bbury=355pt,height=8cm}}
\caption{ \it{ The ratio $R_1 = \fleisher$ as a function of the parameter $\rhobar$ taken from \cite{ref:aligreub},
for $\etabar=0.25$ (lower curve) and $\etabar=0.52$ (upper curve). The horizontal
thick lines show the CLEO measurement (with $\pm 1 \sigma$ errors). The shaded vertical
band corresponds to the $\pm 1 \sigma$ interval for $\rhobar$ obtained in the present analysis.}}
\label{fig:alietal}
\end{center}
\end{figure}

At present, theorists do not agree on the effects of hadronic interactions on this analysis
\cite{ref:bkpi} of the ratio $R_1$. But, considering that these effects are under control, the needed
experimental accuracy on $R_1$ has been evaluated such that this measurement provides
an information on $\rhobar$ of similar precision as the one obtained at present.
The model of \cite{ref:aligreub} has been used in which the authors have studied the variation of $R_1$ with the 
$\rhobar$ parameter (Figure \ref{fig:alietal}). The present determination of $\rhobar$ corresponds to:
\begin{equation}
R_1~=~\fleisher~=0.87 \pm 0.07~ 
\end{equation}

\section{Tests of the internal consistency of the Standard Model for CP violation.}
\label{sec:param}

Four constraints, three measurements and one limit, have been used until now 
to measure the values of the two parameters $\rhobar$ and $\etabar$. It is also possible
to remove, from the construction of the two dimensional probability distribution for $\rhobar$ and $\etabar$, 
the external information on the value of one of the constraint or of another
parameter entering into the Standard Model expressions for the constraints.
The results will have some dependence in the central values taken for all the other parameters
but, the main point in this study, is to compare the uncertainty on a given quantity determined in this way
to its present experimental or theoretical error. This comparison allows to quantify the importance
of present measurements of the different quantities entering into the definition of the allowed region
in the $(\rhobar,~\etabar$) plane.
Results have been summarized in Table \ref{tab:b}.

\subsection{Expected value for the $\Bs-\Bsbar$ oscillation parameter, $\dms$.}

\begin{figure}
\begin{center}
{\epsfig{figure=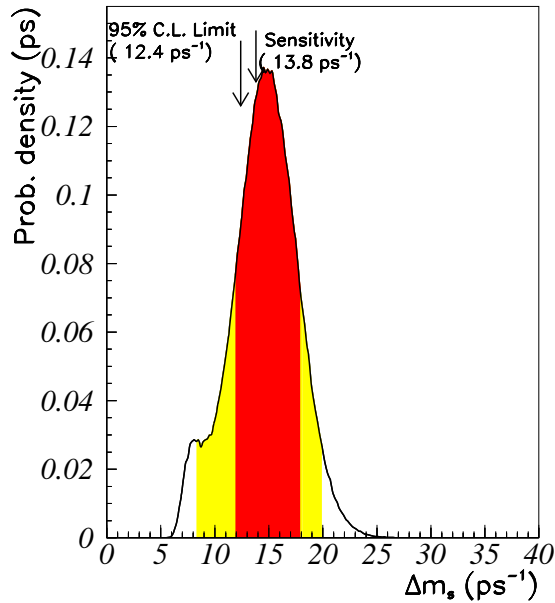,bbllx=0pt,bburx=470pt,bblly=0pt,bbury=515pt,height=8cm}}
\caption{ \it{ The $\Delta m_s$ probability distribution obtained with the same constraints as in Figure \ref{fig:rhoeta_cimento}
 once the constraint from the experimental limit on $\Delta m_s$ has been removed. The dark-shaded and the clear-shaded intervals 
correspond to 68$\%$ and 95 $\%$ confidence level regions respectively. }}
\label{fig:haricot2_dms}
\end{center}
\end{figure}

\begin{table}[htb]
\begin{center}
\begin{tabular}{|c|c|c|c|}
\hline
 parameter           & Results in Scenario I           & Results in Scenario II           &             Present value \\
\hline 
 $\Delta m_s$        & $14.8 ^{+2.8}_{-2.8} ~ps^{-1}$ &  $15.0 ^{+3.3}_{-3.3} ~ps^{-1}$ & $>$ 12.4 ps$^{-1}~{\rm at}~95 \%~C.L.$ \\
  $\vubovcb$         & $ 0.097^{+0.033}_{-0.022}$  &  $0.091^{+0.033}_{-0.024}$     & $0.080 \pm 0.017 / 0.104 \pm 0.019 $   \\
  $B_K$               &  $ 0.87^{+0.34}_{-0.20}$   &  $ 0.90^{+0.34}_{-0.22}$      & $ 0.86 \pm 0.09$    \\
$f_{B_d}\sqrt{B_{B_d}}$ &   ($223 \pm 13$) MeV   & ($228 \pm 14$) MeV        &   $(210 ^{+39}_{-32}) ~MeV$ \\
 $\overline{m_t}(m_t)$  &  $(179^{+52}_{-34})$ GeV    & $(179^{+56}_{-39})$ GeV        & $(167 \pm 5)$ GeV  \\
  $\Vcb$ &   $(42.0^{+8.0}_{-4.0})\times 10^{-3}$   &  $(41.0^{+9.5}_{-4.0})\times 10^{-3} $ &  $(40.4 \pm 1.2)\times 10^{-3}$   \\ 
\hline
\end{tabular}
\caption[]{ \it { Values of the different parameters obtained after having removed, in turn, their contribution into the different
constraints. The results in the two scenarios are reported.}}
\label{tab:b}
\end{center}
\end{table}

Removing the constraint from the measured limit on the mass difference between the strange
B meson mass eigenstates, $\dms$, the density probability distribution for $\dms$
is given in Figure \ref{fig:haricot2_dms}. $\Delta m_s$ is expected to be between [12.0 - 17.6]$ps^{-1}$ within 
one sigma and $<20~ps^{-1}$ at 95$\%$ C.L.
The present limit excludes already a large fraction of this distribution. Present analyses at LEP
are situated in a high probability region for a positive signal and this is still a challenge
for LEP collaborations.

\subsection{Top mass measurement.}

If the information on the top mass measurement by CDF and D0 collaborations is removed,
the value for $\overline{m_t}(m_t)$ is :  $\overline{m_t}(m_t)$ =  $(179^{+52}_{-34})$ GeV.
The present determination of $m_t$ with a $\pm$5 GeV error has thus a large impact for the present analysis.

\subsection{Measurement of $\Vcb$.} 

The central value determined for $\Vcb$ is close to the direct measurement: $\Vcb$=$(42.0^{+8.0}_{-4.0})\times 10^{-3}$.
Direct measurements of $\Vcb$ are thus important to constrain the allowed region in the $(\rhobar,~\etabar$)
plane once their relative error is below 10$\%$. The density probability distribution for the parameter $\Vcb$ is given 
in Figure \ref{fig:haricot2_bk}.

\subsection{Measurement of $\vubovcb$.} 
The central value determined for  $\vubovcb$ is close to the direct measurement:   $\vubovcb$=$ 0.097^{+0.033}_{-0.022}$.
This indirect measurement shows the importance of having a precision on $\vubovcb$ better than 30$\%$. 
The density probability distribution for $\vubovcb$ is given in Figure \ref{fig:haricot2_bk}.

\subsection{Measurement of $B_K$.}

The density distribution for the parameter $B_K$ is given in Figure \ref{fig:haricot2_bk}.
It indicates that:
\begin{itemize}
\item values of $B_K$ smaller than 0.6 are excluded at 98.4$\%$ C.L.,
\item large values of $B_K$ are compatible with the other constraints over a large domain.
\end{itemize}

The present estimate of $B_K$, from lattice QCD, with a 10$\%$ relative error has thus a large
impact for the present analysis.

\subsection {Measurement of $f_{B_d} \sqrt {B_{B_d}}$ }
\label{sec:523}

\begin{figure}
\begin{center}
{\epsfig{figure=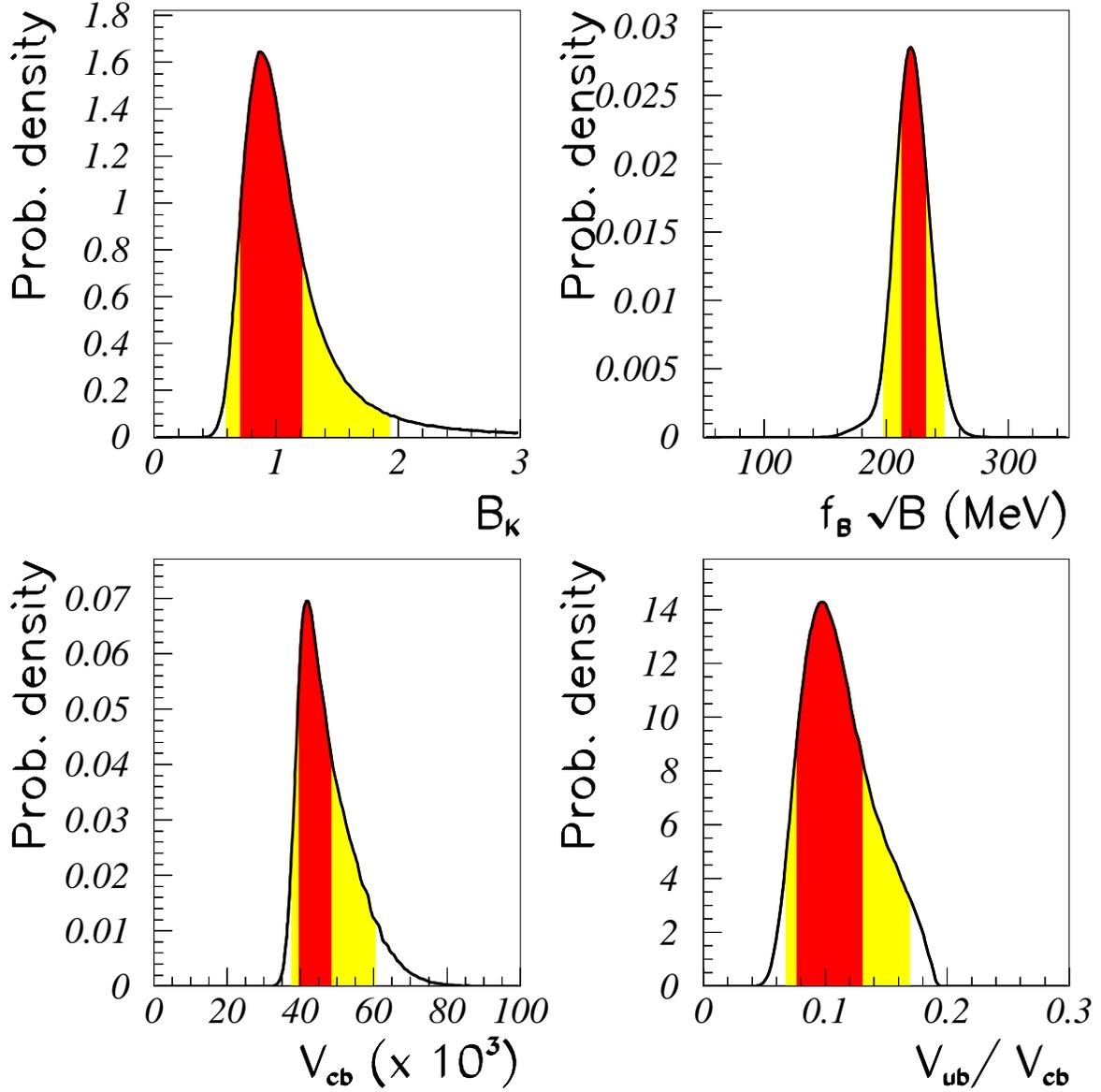,bbllx=0pt,bburx=525pt,bblly=0pt,bbury=525pt,height=16cm}}
\caption{ \it{ The $B_K$, $f_{B_d}\sqrt{B_{B_d}}$, A and $\vubovcb$ probability distributions obtained with the same 
constraints as in Figure \ref{fig:rhoeta_cimento} once the concerned parameter has been removed. 
The dark-shaded and the clear-shaded intervals correspond to 68$\%$ and 95 $\%$ confidence level regions respectively. }}
\label{fig:haricot2_bk}
\end{center}
\end{figure}

A rather accurate value is obtained:
\begin{equation}
f_{B_d}\sqrt{B_{B_d}} = \left  ( 223 \pm 13 \right ) MeV
\label{eq:fb_mes}
\end{equation}
This result is, in practice, in agreement and more precise than the present evaluation 
of this parameter (\ref{eq:fbsqrtb}) which has been determined measuring $f_{D_s}$ and using 
results from lattice QCD on $\frac{f_{B_d}}{f_{D_s}}$. Table \ref{tab:fbpre} gives the contributions of the
uncertainties, on the different parameters, to the error on $ f_{B_d}\sqrt{B_{B_d}}$ extracted from the ratio $f_{B_d}/ f_{D_s} $. 
An effort has to be done to improve both the experimental and the
theoretical precisions. Finally, from the experimental point of view, a $\tau/$Charm factory could provide the ultimate 
precision on $f_{D_s}$ and $f_{D^+}$.
The density probability distribution for the parameter $f_{B_d}\sqrt{B_{B_d}}$ is given in Figure \ref{fig:haricot2_bk}.

\section{Measurement of the angles $\theta$, $\theta_u$, $\theta_d$ and $\phi$.}
\label{sec:masse}

This section is an update of the results presented in \cite{ref:bello}, a similar study can be found also in \cite{ref:barbi2}.
There are nine possibilities to introduce the CP violation phase into the elements of the CKM matrix \cite{ref:friche}.
The authors of \cite{ref:friche} have argued for a parametrization, based on the observed hierarchy in the values 
of quark masses. Several theoretical works \cite{ref:barbierimasse} show that the observed pattern of fermion masses and mixing 
angles could originate from unified theories with an U(2) flavour symmetry.
The authors of \cite{ref:friche} have introduced four angles which have simple physical interpretations. 
$\theta$ corresponds to the mixing between the families 2 and 3. $\theta_{u(d)}$ is the mixing angle between families
1 and 2, in the up(down) sectors. Finally $\phi$ is responsible for CP
violation and appears only in the elements of the C.K.M. matrix relating the first two families.

This parametrization is given below:

\begin{equation}
\begin{array}{ccc}
V_{CKM} =
&
\left ( \begin{array}{cccc}
s_u s_d c+ c_u c_d e^{-i\phi}~~~~~~~~~~s_uc_dc - c_us_d e^{-i\phi} ~~~     s_us ~~\\
c_us_dc - s_uc_d  e^{-i\phi}  ~~~~~~~~~~    c_uc_dc + s_us_d e^{-i\phi}~~~   c_us ~~\\
-s_d s   ~~~~~~~~~~~~~~~~~~~~~~   -c_ds ~~~~~~~~~~~~~~~~~  c
\end{array} \right )
\end{array}
\end{equation}
where $c_x$ and $s_x$ stand for $cos \theta_x$ and  $sin \theta_x$ respectively.

The four angles are related to the modulus of the following C.K.M. elements:
\begin{equation}
\sin{\theta} = \Vcb \sqrt{1 + \vubovcb^2}
\end{equation}
\begin{equation}
\tan{\theta_u} = \vubovcb
\end{equation}
\begin{equation}
\tan{\theta_d} = \vtdovts
\end{equation}
and
\begin{equation}
\phi= \arccos \left ( \frac {\sin^2{\theta_u} \cos^2{\theta_d} \cos^2{\theta} +
\cos^2{\theta_u} \sin^2{\theta_d} -\left | {\rm V}_{us} \right |^2}
{2 \sin{\theta_u}\cos{\theta_u}\sin{\theta_d}\cos{\theta_d}\cos{\theta} } \right )
\label{eq:phi1}
\end{equation}

The first three equations illustrate the direct relation between the angles $\theta$,
$\theta_u$ and $\theta_d$ and the measurements of B decay and oscillation parameters.

The angle $\phi$ has also a nice interpretation because, in the limit where $\theta=0$ (in practice
$\theta \simeq 2^{\circ}$), the elements $ {\rm V}_{us} $ 
and ${\rm V}_{cd} $ have the same modulus, equal to $\sin{\theta_c}$ 
(${\theta_c}$ is the Cabibbo angle) and can be represented in a complex plane
by the sum of two vectors, of respective lengths $\sin{\theta_u} \cos{\theta_d}$
and  $\sin{\theta_d} \cos{\theta_u}$, making a relative angle $\phi$.
It can be shown that this triangle is congruent to the usual unitarity triangle
\cite{ref:friche} and that $\phi \simeq \alpha$.

\begin{table}[htb]
\begin{center}
\begin{tabular}{|c|c|c|}
\hline
 Angle      &         measured value               &   value expected from quark masses \cite{ref:quark_mass}    \\ \hline
$\theta$    &   (2.35  $\pm$ 0.07)$^{\circ}$       &                                                              \\
$\theta_u$  &   (5.00  $\pm$ 0.45)$^{\circ}$       &     (3.4  $\pm$ 0.4)$^{\circ}$                           \\  
$\theta_d$  &   (11.2  $\pm$ 0.7)$^{\circ}$        &     (12.6 $\pm$ 1.2)$^{\circ}$                          \\  
$\phi$      &   (96.7 $\pm$ 8.1)$^{\circ}$         &     (85    $\pm$ 21)$^{\circ}$                            \\
\hline
\end{tabular}
\caption[]{ \it { Values for the angles of the parametrization \cite{ref:friche}, 
compared with those obtained using the values of the quark masses as given in 
\cite{ref:quark_mass} evaluated at Q$^2$=M$_W^2$ ( the values used for the quark masses are: 
$m_u = 2.35^{+0.42}_{-0.45} MeV/c^2,~m_d = 4.73^{+0.61}_{-0.67} MeV/c^2,~m_s = 94.2^{+11.9}_{-13.1} MeV/c^2~and~
m_c = 684^{+56}_{-61} MeV/c^2$ ) }}
\label{tab:tabmass}
\end{center}
\end{table}

\begin{figure}
\begin{center}
{\epsfig{figure=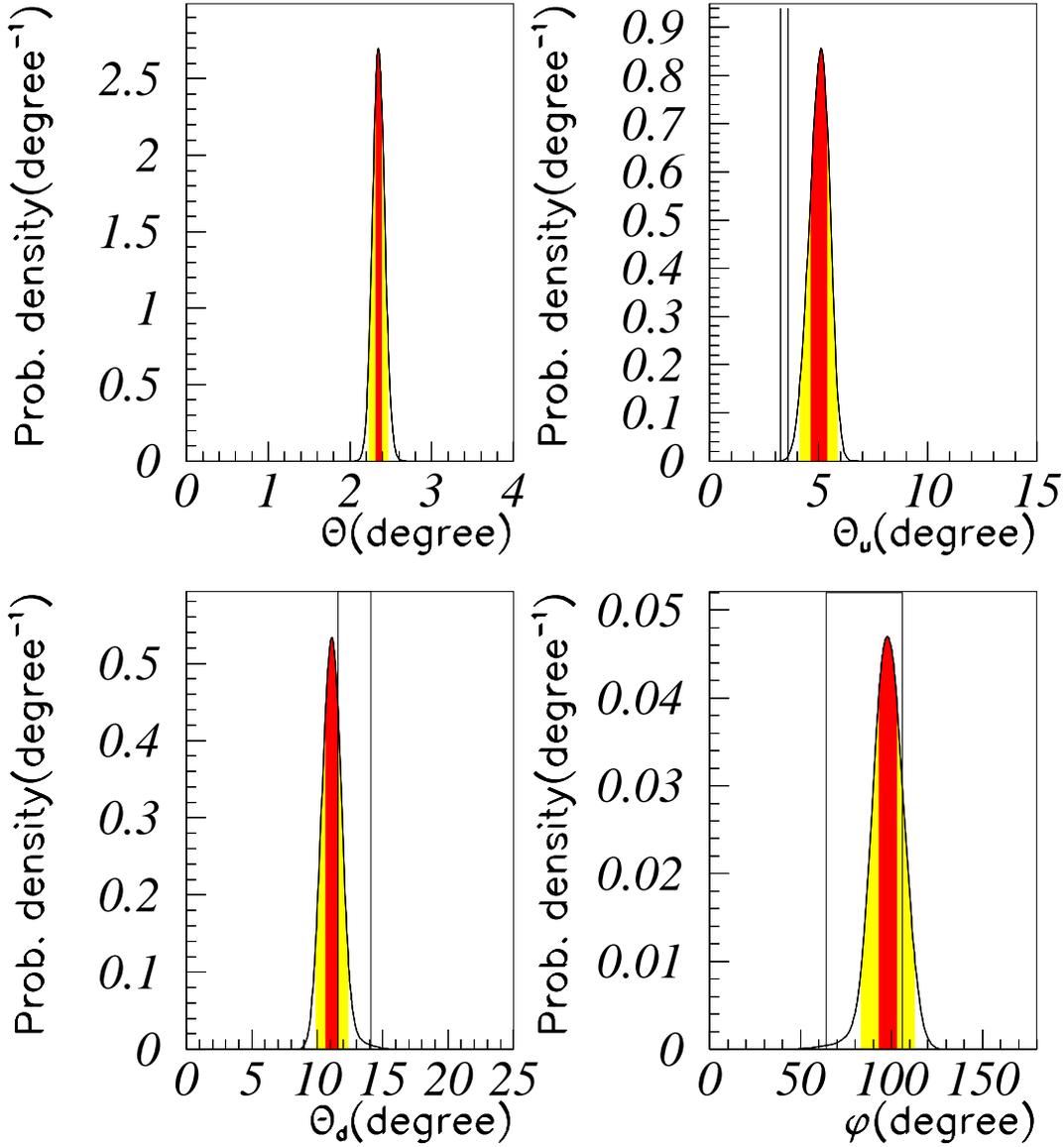,bbllx=0pt,bburx=538pt,bblly=0pt,bbury=519pt,height=16cm}}
\caption{ \it {  The distributions of the angles $\theta$, $\theta_u$, $\theta_d$ and $\phi$ proposed in the 
parametrisation \cite{ref:friche}. The dark-shaded and the clear-shaded intervals correspond to 68$\%$ and 95 $\%$ confidence level
regions respectively. The lines represent the $\pm$1 $\sigma$ region corresponding to the values of the angles obtained using the 
values for quark masses given in \cite{ref:quark_mass}. }}
\label{fig:friche}
\end{center}
\end{figure}

Using the constraints defined previously, the respective probability distributions
for the four angles have been given in Figure \ref{fig:friche}; the numerical values are summarized in Table \ref{tab:tabmass}.

In this parametrization of the C.K.M. matrix the relation between the mixing angles and the quark masses 
can be written as \cite{ref:friche2}:
\begin{equation}
\tan{\theta_u}=\sqrt{\frac{m_u}{m_c}},~\tan{\theta_d}=\sqrt{\frac{m_d}{m_s}}.
\label{eq:phi2}
\end{equation}

Using the values for the quark masses given in \cite{ref:quark_mass} evaluated at Q$^2$=M$_W^2$,
the values for the angles $\theta_u$ and $\theta_d$ are given 
in Table \ref{tab:tabmass}.
In this interpretation, the angle $\phi$ can be obtained using (\ref{eq:phi1}) and (\ref{eq:phi2}).
Present measurements support a value of $\phi$ close to $90^{\circ}$
which corresponds to the maximal CP violation scenario of \cite{ref:friche}.\\
The present analysis indicates that a higher value for the ratio $m_u/m_c$ is favoured or that the expression 
relating $\theta_u$ to the $u$ and $c$ quark masses has to be corrected. 

\section{Conclusions.}
The $\overline{\rho}$ and $\overline{\eta}$ parameters have been determined using the constraints from the measurements of
$  \frac{\left | V_{ub} \right |}{\left | V_{cb} \right |} $, $\epsilonk$, $\Delta m_d$ and from the limit on
$\Delta m_s$:
$$
\rhobar=0.202 ^{+0.053}_{-0.059}   ,~\etabar=0.340 \pm 0.035 
$$
Contrary to similar studies in this field, which claim a rather symmetric interval of variation for $\rhobar$, 
around zero \cite{ref:concurrence}, the negative $\rhobar$ region is excluded at 99.2$\%$ C.L..
The present analysis assumes the unitarity of the CKM matrix. In this framework the values of $sin 2\beta$, $sin 2\alpha$ and
$\gamma$ have been also deduced. They are :
$$
\sin 2 \alpha =  -0.26 ^{+ 0.29}_{-0.28}  ,~\sin 2 \beta =  0.725 ^{+0.050}_{-0.060}   ,~\gamma= (59.5^{+8.5}_{-7.5})^{\circ} 
$$
The value of $ sin 2\beta$ has an accuracy similar to the one expected after 3-4 years of running at B factories. 
An interesting outcome from this study is that a more conservative approach in the choice of the parameters ( Scenario II) gives a 
less than 10$\%$ variation on the uncertainties attached to the various measured quantities.
The internal consistency of the Standard Model expectation for CP violation, expressed by a single
phase parameter in the C.K.M. matrix, has been verified by removing, in turn, the different
constraints imposed by the external parameters. No anomaly has been noticed with respect
to the central values used in the present analysis. This study has mainly quantified the needed accuracy
on the determinations of these parameters so that they bring useful constraints in the
determination of $\rhobar$ and $\etabar$. In this respect, present uncertainties on $m_t$, $\Vcb$,$\vubovcb$ and
$B_K$ have important contributions.
Low values of $B_K$, below 0.6, are not compatible with the present analysis at 98.4$\%$ C.L..
$\Delta m_s$ is expected to lie, with 68$\%$ C.L., between 12.0 and 17.6 $ps^{-1}$. More accurate measurements, still expected at 
CLEO and LEP, and more precise evaluations of non perturbative QCD parameters from lattice QCD, will improve these results in 
the coming years. A Tau/Charm factory providing accurate values for $f_{D^+}$ and $f_{D_s}$ is expected to have important 
contributions in this analysis. \\
Another parametrization of the CKM matrix, proposed in \cite{ref:friche} has been studied.  The four corresponding parameters, 
which are angles, have been determined. The present analysis indicates that a lower value for the ratio $m_u/m_c$ is favoured 
or that the expression relating $\theta_u$ to the $u$ and $c$ quark masses has to be corrected.

\section{Acknowledgments}
We would like to thank C.W. Bernard, A. S. Kronefeld and R. Gupta for all the useful interactions on the subjects related to non 
perturbative QCD parameters evaluation. A special thanks to M. Mazzucato, F. Richard, D. Treille and W. Venus for constant and 
warm support in this work. We would like to thank D. Bloch for the careful reading of this manuscript.
This work includes a large sample of data from various theoretical analyses and experimental results from different 
collaborations but it has been made possible for us only because we are members of the DELPHI Collaboration in which we acquired 
some familiarity in B physics.

\end{document}